\documentclass[a4paper,aps,prd,10pt,preprintnumbers,showpacs,showkeys,twocolumn,superscriptaddress,nofootinbib,amsmath,amssymb]{revtex4-1}
\usepackage{cmap}
\usepackage[utf8]{inputenc}
\usepackage[T1]{fontenc}
\usepackage{graphicx,orcidlink,xcolor,ulem}
\usepackage{soul}

\def\imo{i}
\def\K{{\cal K}}
\def\re#1{Re(#1)}
\def\im#1{Im(#1)}

\def\Order#1{{\cal O}\left(#1\right)}

\begin{document}

\title{Long-lived quasinormal modes, grey-body factors and absorption cross-section of the black hole immersed in the Hernquist galactic halo}

\author{B. C. Lütfüoğlu}
\email{bekir.lutfuoglu@uhk.cz}
\affiliation{Department of Physics, Faculty of Science, University of Hradec Králové, \\
Rokitanskeho 62/26, Hradec Králové, 500 03, Czech Republic. }

\begin{abstract}
We analyze quasinormal modes, grey-body factors, and absorption cross-sections of a massive scalar field in the background of a Schwarzschild black hole surrounded by a Hernquist dark-matter halo. The quasinormal spectrum is obtained through the higher-order WKB method and verified by time-domain evolution, showing consistent results. The field mass increases the oscillation frequency and reduces the damping rate, producing longer-lived modes, while variations in the halo parameters lead to moderate shifts in the spectrum.  The grey-body factors reveal a suppression of low-frequency transmission and a displacement of their main features toward higher frequencies, resulting in a corresponding shift in the absorption cross-section. 
\end{abstract}

\maketitle

\section{Introduction}

Black holes surrounded by galactic dark-matter halos constitute a realistic astrophysical setting where strong gravity and large-scale structure coexist. Observations indicate that both stellar-mass and supermassive black holes reside within extended distributions of dark matter that dominate the galactic mass budget. Although the local dark-matter density near the event horizon is minute compared to the black-hole energy density, even a weakly interacting halo can subtly influence the propagation of perturbative fields and thus modify observable quantities such as quasinormal spectra and Hawking emission rates. In addition, the spectral and other types of instabilities have been extensively discussed in this context. The recent literature on the above topics is extensive and we mention here only some of these works  ~\cite{Liu:2024xcd, Chen:2024lpd, Konoplya:2025nqv, Hamil:2025pte, Pezzella:2024tkf, Konoplya:2022pbc, Becar:2024agj, Liu:2023vno, Konoplya:2023hqb, Jha:2024ltc, Liu:2024bfj, Mollicone:2024lxy, Dubinsky:2025fwv, Cheung:2021bol, Destounis:2025dck, Rosato:2024arw, Courty:2023rxk, Mascher:2022pku}. Therefore, the study of black-hole perturbations in dark-matter environments has become an important extension of classical perturbation theory, linking gravitational-wave phenomenology with the physics of the surrounding medium \cite{Berti:2025hly,Destounis:2023ruj}.

Quasinormal modes (QNMs) are the characteristic damped oscillations describing how a black hole relaxes after a perturbation~\cite{Kokkotas:1999bd, Nollert:1999ji, Bolokhov:2025uxz, Konoplya:2011qq}. Their complex frequencies encode both oscillation and damping rates and carry direct information about the underlying geometry and its perturbative stability. Deviations in QNM spectra from the standard vacuum predictions may thus reveal traces of the environment, alternative gravity effects, or new fields coupled to curvature. For black holes immersed in galactic halos, the halo parameters modify the background metric, leading to detectable shifts in quasinormal frequencies and damping times~\cite{Konoplya:2025nqv,Liu:2024xcd,Chen:2024lpd}.  

An effective and physically transparent approach to model the interaction between a black hole and its environment is through a test scalar field endowed with an effective mass term,
\begin{equation}
\Box \Phi - \mu_{\rm eff}^{2}\Phi = 0,
\end{equation}
where $\mu_{\rm eff}$ represents the mass of the field. Depending on the underlying physics, $\mu_{\rm eff}$ may correspond to  either the mass of a field propagating in the black hole's background or an effective mass corresponding, for example, to the background magnetic field~\cite{Konoplya:2008hj, Chen:2011jgd, Konoplya:2007yy, Kokkotas:2010zd, Davlataliev:2024mjl}. The introduction of this term alters the asymptotic structure of the effective potential, often producing quasi-bound states or long-lived modes known as {\it quasi-resonances}~\cite{Ohashi:2004wr, Konoplya:2017tvu}. These long-lived oscillations can substantially affect not only the ringdown frequencies but also the time-domain evolution of perturbations and modify the ringdown signal of gravitational waves. In addition, the quasi-bound states take place when the mass of the field is turned on. Extensive literature on the above questions can be found in~\cite{Jing:2004zb, Rogatko:2007zz, Konoplya:2006gq, Konoplya:2018qov, Koyama:2001ee, Koyama:2001qw, Konoplya:2002wt, Moderski:2001tk, Konoplya:2005hr, Churilova:2019qph, Dubinsky:2024jqi,Destounis:2019hca,Vlachos:2021weq,Chatzifotis:2021pak,Vieira:2021ozg,Steinhauer:2025bbs,Vieira:2025gcy,Vieira:2025ljl,Vieira:2023ylz,Dubinsky:2025wns,Malik:2025czt,Konoplya:2006br}. In addition, massive fields may play a role in ultra-low-frequency radiation accessible to Pulsar Timing Arrays~\cite{Konoplya:2023fmh}, and appear naturally as effective degrees of freedom in brane-world scenarios, where the bulk-induced tidal term acts as a mass contribution~\cite{Seahra:2004fg, Ishihara:2008re}.

From the theoretical standpoint, the presence of a massive field provides a mechanism for energy exchange between the black hole and its surroundings. If the effective potential allows partial wave trapping, the system may develop superradiance and superradiant instabilities \cite{Starobinskil:1974nkd,Starobinskii:1973vzb,Brito:2015oca}, whose onset depends on both the mass of the field and the parameters describing the dark-matter halo. Consequently, analyzing the dependence of QNMs on these parameters provides insight into the dynamical stability of the configuration and into possible observational signatures of dark matter or external fields. Detailed studies of perturbations and QNMs of massive fields in various modified or dark-matter–inspired geometries can be found in~\cite{Lutfuoglu:2025hjy, Skvortsova:2025cah, Zinhailo:2024kbq, Konoplya:2024wds, Bolokhov:2023ruj, Lutfuoglu:2025eik, Malik:2025ava, Skvortsova:2024eqi, Bolokhov:2024ixe, Fernandes:2021qvr, Lutfuoglu:2025qkt, Konoplya:2007zx, Bolokhov:2023bwm, Zinhailo:2024jzt, Dubinsky:2024hmn, Konoplya:2024ptj, Lutfuoglu:2025bsf, Bolokhov:2024bke, Percival:2020skc}.  

Beyond the dynamical response, another key characteristic of black-hole perturbations is encoded in the grey-body factors (GBFs), which describe the probability that Hawking radiation quanta escape to infinity instead of being reflected by the effective potential barrier. GBFs determine the energy and particle emission rates and thus directly influence the evaporation process~\cite{Page:1976df, Page:1976ki, Kanti:2002nr}. 
Interestingly, GBFs are much more stable to deformations of the background, they can be used in order to describe the ringdown signal, and even construct the signal through destabilized QNM spectra that appear from modified black holes by astrophysical environments \cite{Oshita:2023cjz,Okabayashi:2024qbz,Oshita:2024fzf,Oshita:2025ibu,Oshita:2024wgt,Rosato:2024arw,Konoplya:2025ixm}. Within the context of gauge/gravity duality, they also have a holographic interpretation: the poles of the transmission coefficient correspond to those of retarded Green’s functions in the dual field theory, linking black-hole scattering to transport properties such as optical conductivity in strongly coupled plasmas or holographic superconductors~\cite{Horowitz:2008bn, Herzog:2009xv, Konoplya:2009hv}.  

In \cite{Konoplya:2021ube, Cardoso:2021wlq} QNMs and GBFs were found for massless fields in the background of the black hole immersed in a dark-matter halo described by the Hernquist halo~\cite{Hernquist:1990be}. Here, we extend that study to the analysis of the spectrum and GBFs for a massive field, where the effective mass term could be interpreted as a contribution from the surrounding magnetic field. The analysis of the spectral and scattering properties of massive fields in the black hole against variations of the surrounding dark-matter environment reveals richer physics in both the time and frequency domains.

The structure of the paper is as follows. In Sec.~\ref{sec:background} we present the analytic form and geometric properties of the Schwarzschild black hole embedded in the Hernquist dark-matter halo. Section~\ref{sec:methods} describes the numerical and semi-analytic methods employed to determine quasinormal frequencies, including the higher-order WKB approach with Padé approximants and the time-domain integration technique with Prony analysis. In Sec.~\ref{Quasinormal modes} we discuss the obtained spectra, their dependence on the field mass and halo parameters, and the consistency between different computational methods. Section~\ref{sec:GBF} is devoted to GBFs and energy transmission, followed by a summary of the main conclusions in Sec.~\ref{sec:conc}.

\section{Black Hole metric and the effective potential}\label{sec:background}

\subsection{Black hole solution}

Following the analysis developed in~\cite{Cardoso:2021wlq}, the gravitational background generated by a galactic halo can be effectively described by the Hernquist density profile~\cite{Hernquist:1990be}, which has been successfully employed to reproduce the observed S\'ersic surface-brightness distributions of elliptical galaxies and galactic bulges. The matter density is modeled as
\begin{equation}
\rho(r)=\frac{M\,a_{0}}{2\pi\,r\,(r+a_{0})^{3}},
\label{eq:Hernquist_density}
\end{equation}
where $M$ represents the total mass attributed to the halo, while $a_{0}$ characterizes the typical radial scale of the system. For $r\gg a_{0}$ the profile decays as $\rho\propto r^{-4}$, ensuring a finite total mass, whereas at small radii the density follows $\rho\propto r^{-1}$, providing a realistic central cusp consistent with numerical simulations of galactic halos.

The static and spherically symmetric geometry generated by this distribution can be expressed in Schwarzschild-like coordinates as
\begin{equation}
ds^{2}=-f(r)\,dt^{2}+\frac{dr^{2}}{1-\dfrac{2m(r)}{r}}+r^{2}d\Omega^{2},
\label{eq:metric_form}
\end{equation}
where $m(r)$ is the mass function that encodes both the contribution of the central black hole and the extended halo. A convenient, though not unique, parametrization of $m(r)$ that reproduces the Hernquist distribution~\eqref{eq:Hernquist_density} and ensures asymptotic flatness is
\begin{equation}
m(r)=M_{\mathrm{BH}}+\frac{M\,r^{2}}{(r+a_{0})^{2}}\!\left(1-\frac{2M_{\mathrm{BH}}}{r}\right)^{2}.
\label{eq:mass_function}
\end{equation}
At short distances, the term dominated by $M_{\mathrm{BH}}$ describes a black hole of mass $M_{\mathrm{BH}}$, while for $r\gg a_{0}$ the mass function approaches $M_{\mathrm{BH}}+M$, corresponding to the total (ADM) mass of the system.

When this matter distribution is used as the source of the Einstein equations, the temporal metric component $f(r)$ obtained in~\cite{Cardoso:2021wlq} takes the form
\begin{align}
f(r)&=\!\left(1-\frac{2M_{\mathrm{BH}}}{r}\right)\exp(\Upsilon), \label{eq:f_metric}\\[2mm]
\Upsilon&=-\pi\!\sqrt{\frac{M}{\xi}}
+2\!\sqrt{\frac{M}{\xi}}\,
\arctan\!\!\left(\frac{r+a_{0}-M}{\sqrt{M\,\xi}}\right),\\[1mm]
\xi&=2a_{0}-M+4M_{\mathrm{BH}}.
\end{align}
This expression satisfies the correct asymptotic limit $f(r)\to 1$ for $r\to\infty$ and thus represents an asymptotically flat spacetime. The physical event horizon is located at $r_{h}=2M_{\mathrm{BH}}$, while the total mass measured at infinity equals $M_{\mathrm{ADM}}=M_{\mathrm{BH}}+M$.

The corresponding matter density inferred from the Einstein equations can be obtained via
\begin{equation}
4\pi\rho(r)=\frac{1}{r^{2}}\frac{d m(r)}{dr}
   =\frac{2M(a_{0}+2M_{\mathrm{BH}})}
          {r\,(r+a_{0})^{3}}\left(1-\frac{2M_{\mathrm{BH}}}{r}\right).
\label{eq:Hernquist_density_GR}
\end{equation}
At large $r$ and for galaxies where $a_{0}\gg M_{\mathrm{BH}}$, this expression asymptotically reproduces the standard Hernquist profile~(\ref{eq:Hernquist_density}), validating the interpretation of $M$ as the halo mass. In observationally relevant systems, the typical scaling relation is $a_{0}\gtrsim10^{4}M$, which ensures that the dark-matter contribution dominates only far from the black-hole horizon.

In what follows, we shall focus on the astrophysically motivated parameter domain
\[
M_{\mathrm{BH}}\ll M\ll a_{0},
\]
where the central object represents a relatively small compact source embedded within a massive but diffuse Hernquist halo. This hierarchy allows one to explore how the dark-matter environment modifies QNM spectra and GBFs of perturbative fields without strongly altering the near-horizon geometry of the black hole. 

QNMs and GBFs of massless fields for this solution have been analyzed in \cite{Konoplya:2021ube}, while various generalizations for the background metric for more general profiles of halo were found and studied in \cite{Konoplya:2022hbl, Figueiredo:2023gas, Speeney:2024mas, Pezzella:2024tkf, Fernandes:2025osu, Destounis:2025tjn, Jusufi:2022jxu, Konoplya:2025ect}.

\subsection{Perturbations of a massive scalar field}

Small disturbances of a black hole by external fields excite a set of damped oscillations known as QNMs. These oscillations describe the characteristic relaxation of the spacetime back to equilibrium through the emission of gravitational or scalar radiation. Mathematically, QNMs are the eigenmodes of the linearized perturbation equations subject to boundary conditions that allow energy to flow into the event horizon and radiate away to infinity.

For a static, spherically symmetric geometry, the dynamics of a minimally coupled scalar field can be separated into radial and angular parts according to
\begin{equation}
\Phi(t,r,\theta,\phi)
=\sum_{\ell,m}\frac{\Psi_{\ell m}(r)}{r}\,
Y_{\ell m}(\theta,\phi)\,e^{-i\omega t},
\label{eq:decomposition}
\end{equation}
which leads to a one-dimensional wave-like equation of Schrödinger type,
\begin{equation}
\frac{d^{2}\Psi_{\ell}}{dr_{*}^{2}}
+\!\left[\omega^{2}-V_{\ell}(r)\right]\Psi_{\ell}=0.
\label{eq:wave_equation}
\end{equation}
Here, the tortoise coordinate $r_{*}$ is defined through the differential relation $dr_{*}/dr=1/f(r)$, ensuring that the entire exterior region of the black hole is mapped onto $r_{*}\in(-\infty,+\infty)$. The effective potential governing scalar perturbations of mass $\mu$ takes the form
\begin{equation}
V(r)=f(r)\!\left(\frac{\ell(\ell+1)}{r^{2}}+\frac{f'(r)}{r} + \mu^{2}\right).
\label{eq:effective_potential}
\end{equation}

In contrast with the case of a massless field, the potential~\eqref{eq:effective_potential} tends to a constant value at spatial infinity, $\lim_{r\rightarrow\infty}V(r)=\mu^{2},$ so that the asymptotic behavior of the solutions depends explicitly on the relation between $\omega$ and $\mu$. The effective mass term can be ascribed to an external magnetic field in the black hole background as shown, for example, in \cite{Konoplya:2007yy,Wald:1974np,Davlataliev:2024mjl}.

\begin{figure}
\resizebox{\linewidth}{!}{\includegraphics{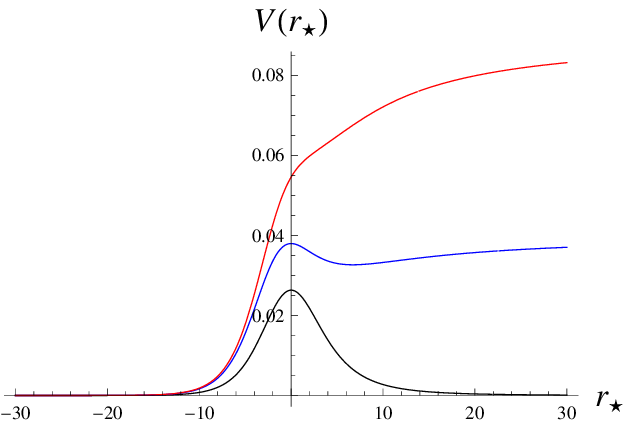}}
\caption{Effective potentials for $\ell=0$ perturbations at $M_{BH}=1$, $M=10 M_{BH}$ $a_{0}=0.1$, $\mu=0$ (black), $\mu=0.2$ (blue) and $\mu=0.3$ (red).}\label{fig:potential1}
\end{figure}

\begin{figure}
\resizebox{\linewidth}{!}{\includegraphics{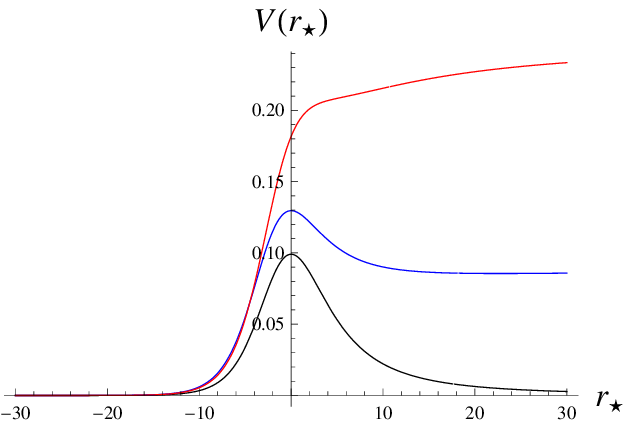}}
\caption{Effective potentials for $\ell=1$  perturbations at $M_{BH}=1$, $M=10 M_{BH}$ $a_{0}=0.1$, $\mu=0$ (black), $\mu=0.3$ (blue) and $\mu=0.5$ (red).}\label{fig:potential2}
\end{figure}

\begin{figure}
\resizebox{\linewidth}{!}{\includegraphics{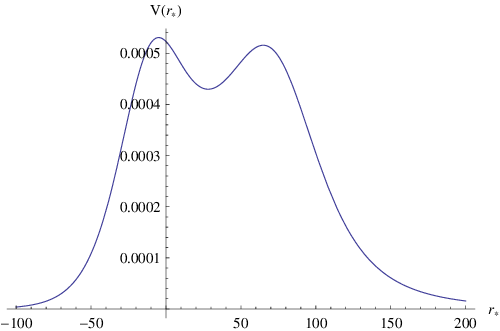}}
\caption{Effective potentials for $\ell=0$  perturbations at $M_{BH}=1$, $M=8 M_{BH}$, $a_{0}=5$, $\mu=0$. The double peak potential does not appear in the regime of realistic galactic halos.}\label{fig:potential3}
\end{figure}

When the mass of the field becomes sufficiently large, the potential peak gradually becomes smoother and may eventually disappear (see   Figs.~\ref{fig:potential1} and~\ref{fig:potential2}). Notice that for unrealistically dense halos with $M \sim a_{0}$, but still corresponding to a black hole solution, the effective potential of even massless perturbations develops a double–peak structure (see, for instance, Fig.~\ref{fig:potential3}). As a result, strong echoes are expected to appear. Since such echoes are clearly not observed, this provides an additional reason for excluding the regime of exceptionally dense halos from our analysis.

The appropriate quasinormal boundary conditions are imposed in terms of the tortoise coordinate:
\begin{equation}
\Psi_{\ell}(r_{*})\sim
\begin{cases}
e^{-i\omega r_{*}}, & r_{*}\to -\infty,\\[6pt]
e^{+i\Omega r_{*}}, & r_{*}\to +\infty,
\end{cases}
\label{eq:BC_massive}
\end{equation}
where $\Omega=\sqrt{\omega^{2}-\mu^{2}}$ such that $\re{\omega}$ and $\re{\Omega}$ have the same sign \cite{Konoplya:2004wg}, so that no incoming radiation is allowed from infinity. The resulting QNM spectrum is a discrete set of complex frequencies,
\[
\omega=\omega_{\mathrm{R}}-i\,\omega_{\mathrm{I}},
\]
where the real part $\omega_{\mathrm{R}}$ determines the oscillation frequency, and the imaginary part $\omega_{\mathrm{I}}>0$ quantifies the damping rate.

\begin{table}
\begin{tabular}{c l c c l}
\hline
$a_{0}$ & $\mu$  & WKB-6, $m=3$ & WKB-7, $m=3$ &  diff. $(\%)$ \\
\hline
$100$ & $0$ & $0.101146-0.094494 i$ & $0.101134-0.094354 i$ & $0.101\%$\\
$100$ & $0.01$ & $0.101166-0.094420 i$ & $0.101152-0.094277 i$ & $0.104\%$\\
$100$ & $0.05$ & $0.101617-0.092647 i$ & $0.101561-0.092448 i$ & $0.151\%$\\
$100$ & $0.1$ & $0.102816-0.087104 i$ & $0.102876-0.087198 i$ & $0.0823\%$\\
$100$ & $0.15$ & $0.103682-0.079082 i$ & $0.103425-0.078781 i$ & $0.304\%$\\
$10^3$ & $0$ & $0.110834-0.103538 i$ & $0.110821-0.103386 i$ & $0.100\%$\\
$10^3$ & $0.01$ & $0.110856-0.103456 i$ & $0.110842-0.103300 i$ & $0.103\%$\\
$10^3$ & $0.05$ & $0.111353-0.101505 i$ & $0.111292-0.101286 i$ & $0.151\%$\\
$10^3$ & $0.1$ & $0.112615-0.095470 i$ & $0.112627-0.095488 i$ & $0.0146\%$\\
$10^3$ & $0.15$ & $0.113739-0.086576 i$ & $0.113242-0.085781 i$ & $0.656\%$\\
$10^4$ & $0$ & $0.111838-0.104475 i$ & $0.111825-0.104322 i$ & $0.100\%$\\
$10^4$ & $0.01$ & $0.111860-0.104393 i$ & $0.111845-0.104236 i$ & $0.103\%$\\
$10^4$ & $0.05$ & $0.112361-0.102424 i$ & $0.112300-0.102203 i$ & $0.151\%$\\
$10^4$ & $0.1$ & $0.113634-0.096335 i$ & $0.113646-0.096351 i$ & $0.0137\%$\\
$10^4$ & $0.15$ & $0.114771-0.087359 i$ & $0.114267-0.086550 i$ & $0.661\%$\\
$10^5$ & $0$ & $0.111938-0.104569 i$ & $0.111925-0.104416 i$ & $0.100\%$\\
$10^5$ & $0.01$ & $0.111960-0.104487 i$ & $0.111946-0.104330 i$ & $0.103\%$\\
$10^5$ & $0.05$ & $0.112463-0.102516 i$ & $0.112401-0.102295 i$ & $0.151\%$\\
$10^5$ & $0.1$ & $0.113736-0.096422 i$ & $0.113748-0.096438 i$ & $0.0137\%$\\
$10^5$ & $0.15$ & $0.114874-0.087438 i$ & $0.114370-0.086628 i$ & $0.661\%$\\
\hline
\end{tabular}
\caption{QNMs of the black hole immersed in galactic halo ($M=10 M_{BH}$, $M_{BH}=1$, $\ell=0$, $n=0$) calculated using the WKB formula at various orders.}
\end{table}

\begin{table}
\begin{tabular}{c l c c l}
\hline
$a_{0}$ & $\mu$  & WKB-6, $m=3$ & WKB-7, $m=3$ &  diff. $(\%)$ \\
\hline
$100$ & $0$ & $0.264885-0.088109 i$ & $0.264887-0.088108 i$ & $0.0009\%$\\
$100$ & $0.01$ & $0.264925-0.088085 i$ & $0.264928-0.088084 i$ & $0.0009\%$\\
$100$ & $0.1$ & $0.268936-0.085654 i$ & $0.268937-0.085654 i$ & $0.0006\%$\\
$100$ & $0.2$ & $0.281187-0.078054 i$ & $0.281176-0.078054 i$ & $0.0038\%$\\
$100$ & $0.3$ & $0.301824-0.064584 i$ & $0.301675-0.064457 i$ & $0.0632\%$\\
$100$ & $0.4$ & $0.330376-0.044188 i$ & $0.330432-0.043250 i$ & $0.282\%$\\
$10^3$ & $0$ & $0.290015-0.096686 i$ & $0.290017-0.096685 i$ & $0.0008\%$\\
$10^3$ & $0.01$ & $0.290059-0.096659 i$ & $0.290062-0.096658 i$ & $0.0008\%$\\
$10^3$ & $0.1$ & $0.294450-0.094009 i$ & $0.294452-0.094009 i$ & $0.0007\%$\\
$10^3$ & $0.2$ & $0.307866-0.085726 i$ & $0.307859-0.085727 i$ & $0.0023\%$\\
$10^3$ & $0.3$ & $0.330517-0.071019 i$ & $0.330426-0.070905 i$ & $0.0431\%$\\
$10^3$ & $0.4$ & $0.361741-0.048304 i$ & $0.361916-0.047856 i$ & $0.132\%$\\
$10^4$ & $0$ & $0.292639-0.097563 i$ & $0.292641-0.097562 i$ & $0.0008\%$\\
$10^4$ & $0.01$ & $0.292683-0.097536 i$ & $0.292686-0.097535 i$ & $0.0008\%$\\
$10^4$ & $0.1$ & $0.297114-0.094862 i$ & $0.297116-0.094862 i$ & $0.0007\%$\\
$10^4$ & $0.2$ & $0.310651-0.086504 i$ & $0.310644-0.086505 i$ & $0.0023\%$\\
$10^4$ & $0.3$ & $0.333508-0.071665 i$ & $0.333417-0.071550 i$ & $0.0428\%$\\
$10^4$ & $0.4$ & $0.365015-0.048741 i$ & $0.365191-0.048294 i$ & $0.130\%$\\
$10^5$ & $0$ & $0.292902-0.097650 i$ & $0.292905-0.097650 i$ & $0.0008\%$\\
$10^5$ & $0.01$ & $0.292947-0.097624 i$ & $0.292949-0.097623 i$ & $0.0008\%$\\
$10^5$ & $0.1$ & $0.297382-0.094948 i$ & $0.297384-0.094947 i$ & $0.0007\%$\\
$10^5$ & $0.2$ & $0.310931-0.086582 i$ & $0.310924-0.086583 i$ & $0.0023\%$\\
$10^5$ & $0.3$ & $0.333808-0.071729 i$ & $0.333717-0.071615 i$ & $0.0428\%$\\
$10^5$ & $0.4$ & $0.365344-0.048784 i$ & $0.365520-0.048337 i$ & $0.130\%$\\
\hline
\end{tabular}
\caption{QNMs of the black hole immersed in galactic halo ($M=10 M_{BH}$, $M_{BH}=1$, $\ell=1$, $n=0$) calculated using the WKB formula at various orders.}
\end{table}

\begin{table}
\begin{tabular}{c l c c l}
\hline
$a_{0}$ & $\mu$  & WKB-6, $m=3$ & WKB-7, $m=3$ &  diff. $(\%)$ \\
\hline
$100$ & $0$ & $0.437398-0.087279 i$ & $0.437399-0.087278 i$ & $0.00015\%$\\
$100$ & $0.01$ & $0.437427-0.087269 i$ & $0.437427-0.087268 i$ & $0.00015\%$\\
$100$ & $0.1$ & $0.440259-0.086293 i$ & $0.440260-0.086293 i$ & $0.00014\%$\\
$100$ & $0.2$ & $0.448882-0.083305 i$ & $0.448882-0.083305 i$ & $0.00011\%$\\
$100$ & $0.3$ & $0.463388-0.078213 i$ & $0.463388-0.078213 i$ & $0.00004\%$\\
$100$ & $0.4$ & $0.483995-0.070814 i$ & $0.483990-0.070815 i$ & $0.00087\%$\\
$100$ & $0.5$ & $0.511003-0.060737 i$ & $0.510996-0.060728 i$ & $0.00218\%$\\
$100$ & $0.6$ & $0.544758-0.047241 i$ & $0.544767-0.047237 i$ & $0.00183\%$\\
$100$ & $0.7$ & $0.589804-0.029028 i$ & $0.583273-0.029126 i$ & $1.11\%$\\
$10^3$ & $0$ & $0.478829-0.095793 i$ & $0.478829-0.095793 i$ & $0.00014\%$\\
$10^3$ & $0.01$ & $0.478860-0.095782 i$ & $0.478861-0.095782 i$ & $0.00014\%$\\
$10^3$ & $0.1$ & $0.481957-0.094720 i$ & $0.481958-0.094719 i$ & $0.00014\%$\\
$10^3$ & $0.2$ & $0.491386-0.091467 i$ & $0.491386-0.091466 i$ & $0.00011\%$\\
$10^3.$ & $0.3$ & $0.507246-0.085927 i$ & $0.507247-0.085927 i$ & $0\%$\\
$10^3$ & $0.4$ & $0.529776-0.077888 i$ & $0.529774-0.077888 i$ & $0.00041\%$\\
$10^3$ & $0.5$ & $0.559318-0.066956 i$ & $0.559312-0.066945 i$ & $0.00217\%$\\
$10^3$ & $0.6$ & $0.596270-0.052341 i$ & $0.596270-0.052341 i$ & $0\%$\\
$10^3$ & $0.7$ & $0.642647-0.033963 i$ & $0.640131-0.032219 i$ & $0.476\%$\\
$10^4$ & $0$ & $0.483160-0.096662 i$ & $0.483160-0.096662 i$ & $0.00014\%$\\
$10^4$ & $0.01$ & $0.483191-0.096651 i$ & $0.483192-0.096651 i$ & $0.00014\%$\\
$10^4$ & $0.1$ & $0.486317-0.095579 i$ & $0.486317-0.095579 i$ & $0.00014\%$\\
$10^4$ & $0.2$ & $0.495830-0.092297 i$ & $0.495830-0.092297 i$ & $0.00011\%$\\
$10^4$ & $0.3$ & $0.511834-0.086708 i$ & $0.511834-0.086708 i$ & $0\%$\\
$10^4$ & $0.4$ & $0.534567-0.078596 i$ & $0.534565-0.078597 i$ & $0.00040\%$\\
$10^4$ & $0.5$ & $0.564376-0.067566 i$ & $0.564370-0.067556 i$ & $0.00217\%$\\
$10^4$ & $0.6$ & $0.601663-0.052821 i$ & $0.601663-0.052821 i$ & $0.00003\%$\\
$10^4$ & $0.7$ & $0.648432-0.034280 i$ & $0.645933-0.032521 i$ & $0.471\%$\\
$10^5$ & $0$ & $0.483595-0.096749 i$ & $0.483595-0.096749 i$ & $0.00014\%$\\
$10^5$ & $0.01$ & $0.483626-0.096738 i$ & $0.483627-0.096738 i$ & $0.00014\%$\\
$10^5$ & $0.1$ & $0.486754-0.095665 i$ & $0.486755-0.095665 i$ & $0.00014\%$\\
$10^5$ & $0.2$ & $0.496276-0.092380 i$ & $0.496277-0.092380 i$ & $0.00011\%$\\
$10^5$ & $0.3$ & $0.512295-0.086786 i$ & $0.512295-0.086786 i$ & $0\%$\\
$10^5$ & $0.4$ & $0.535048-0.078667 i$ & $0.535046-0.078667 i$ & $0.00040\%$\\
$10^5$ & $0.5$ & $0.564884-0.067627 i$ & $0.564878-0.067617 i$ & $0.00217\%$\\
$10^5$ & $0.6$ & $0.602205-0.052868 i$ & $0.602204-0.052868 i$ & $0.00003\%$\\
$10^5$ & $0.7$ & $0.649015-0.034311 i$ & $0.646515-0.032550 i$ & $0.471\%$\\
\hline
\end{tabular}
\caption{QNMs of the black hole immersed in galactic halo ($M=10 M_{BH}$, $M_{BH}=1$, $\ell=2$, $n=0$) calculated using the 6th order WKB formula at various orders.}
\end{table}

\begin{figure}
\resizebox{\linewidth}{!}{\includegraphics{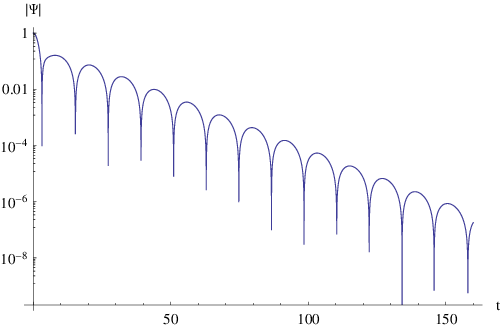}}
\caption{Semi-logarithmic time-domain profile for $\ell=1$, $M_{BH}=1$, $M=10 M_{BH}$, $a=100$, $\mu=0$. The Prony method allows one to find the fundamental QNM $\omega = 0.264892 - 0.0881029 i$, which is very close to the WKB value $\omega =  0.264887 -0.088108 i$.}\label{fig:TD1}
\end{figure}

\begin{figure}
\resizebox{\linewidth}{!}{\includegraphics{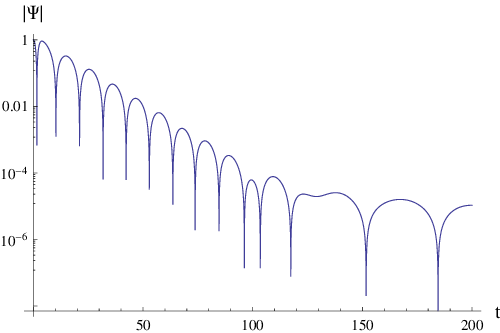}}
\caption{Semi-logarithmic time-domain profile for $\ell=1$, $M_{BH}=1$, $M=10 M_{BH}$, $a=10^5$, $\mu=0.1$. The Prony method allows one to find the fundamental QNM $\omega = 0.297391 - 0.0949604 i$, which is very close to the WKB value $\omega =  0.297382 - 0.094948  i$.}\label{fig:TD2}
\end{figure}

\section{Methods for Computing QNMs}
\label{sec:methods}

\subsection{Time-domain evolution and frequency extraction}

A direct and conceptually transparent way to study the dynamics of perturbations is through numerical evolution in the time domain. Unlike frequency-domain methods, which target stationary solutions of the wave equation, the time-domain approach reveals the complete temporal behavior of the perturbation, from the initial burst to the final decay. It is particularly suitable for analyzing massive fields, multi-barrier potentials, or regimes where the quasinormal signal competes with long-lived tails.

Starting from the master wave equation~\eqref{eq:wave_equation}, we introduce light-cone coordinates
\begin{equation}
u = t - r_*, 
\qquad 
v = t + r_*,
\end{equation}
which transforms the equation into the characteristic form
\begin{equation}
4\,\frac{\partial^2 \Psi}{\partial u\,\partial v}
+V(r)\,\Psi=0.
\label{eq:char_wave}
\end{equation}
This equation is integrated numerically on a null grid using the finite-difference algorithm of Gundlach, Price, and Pullin~\cite{Gundlach:1993tp}. The field value at the future grid point $N$ (with coordinates $u+\Delta, v+\Delta$) is obtained from three previously known points,
\begin{equation}
\Psi_N=\Psi_W+\Psi_E-\Psi_S
-\frac{\Delta^{2}}{8}\,V_S(\Psi_W+\Psi_E)
+\mathcal{O}(\Delta^{4}),
\label{eq:finite_diff}
\end{equation}
where $S$, $E$, and $W$ denote the grid points $(u,v)$, $(u,v+\Delta)$, and $(u+\Delta,v)$, respectively, and $\Delta$ is the step size. A Gaussian pulse is typically used as the initial profile for the field.

The signal $\Psi(t,r)$ recorded at a fixed radius exhibits three distinct regimes: an initial transient depending on the chosen initial data, an exponentially damped ringing governed by the QNMs, and a late-time power-law or exponential tail. The quasinormal ringing stage is the superposition of the dominant fundamental mode and an infinite tower of overtones. To extract the complex frequencies from the intermediate regime, we apply the {\it Prony method}~\cite{Berti:2005ys,Berti:2007dg,Konoplya:2011qq}, which assumes that the waveform can be approximated as a finite sum of exponentially damped oscillations:
\begin{equation}
\Psi(t)=\sum_{n=1}^{N}C_n\,e^{-i\omega_n t}.
\label{eq:prony}
\end{equation}
The amplitudes $C_n$ and the complex frequencies $\omega_n$ are determined from a system of linear prediction equations constructed from consecutive samples of the numerical signal. Provided that the fitting window is dominated by the quasinormal ringing, the Prony procedure yields accurate values for the dominant modes, which can be directly compared with those obtained by semi-analytic methods.

The time-domain evolution is especially valuable when the effective potential develops multiple extrema or when quasi-bound states appear for massive fields, since, in such cases, frequency-domain approaches may converge poorly. In addition, the numerical signal allows one to visualize the entire relaxation process and to assess the stability of the background spacetime against small perturbations. Extensive applications of this approach to a variety of black-hole configurations can be found in~\cite{Malik:2024nhy, Konoplya:2013sba, Dubinsky:2024aeu, Skvortsova:2023zca, Konoplya:2020jgt, Malik:2024elk, Momennia:2022tug, Aneesh:2018hlp, Malik:2024tuf, Qian:2022kaq, Konoplya:2024lch, Churilova:2021tgn, Dubinsky:2024gwo, Dubinsky:2024mwd, Malik:2024qsz, Cuyubamba:2016cug, Stuchlik:2025mjj}.

\subsection{WKB approach and Padé resummation}

To complement the time-domain analysis and obtain semi-analytic estimates of the spectrum, we also employ the higher-order WKB approximation combined with Padé rational expansions. The WKB technique provides an efficient way to determine both quasinormal frequencies and transmission coefficients for smooth, single-barrier potentials of the type encountered in spherically symmetric black-hole perturbations~\cite{Lutfuoglu:2025ljm, Skvortsova:2024atk, Bolokhov:2022rqv, del-Corral:2022kbk, Konoplya:2001ji, Lutfuoglu:2025pzi, Skvortsova:2023zmj, Momennia:2022tug, Zhao:2022gxl, Skvortsova:2024wly, Kodama:2009bf, Lutfuoglu:2025blw, Bolokhov:2025egl}.

Near the maximum of the effective potential, located at $r=r_0$, the WKB quantization rule for complex frequencies reads
\begin{equation}
i\,\frac{\omega^{2}-V_{0}}{\sqrt{-2V_{0}^{\prime\prime}}}
-\sum_{k=2}^{N}\Lambda_{k}
=n+\frac{1}{2},
\label{eq:wkb_condition}
\end{equation}
where $V_{0}$ and $V_{0}^{\prime\prime}$ denote the value and the second derivative of $V(r)$ at the peak, $n$ is the overtone number, and $\Lambda_{k}$ are the higher-order correction terms given explicitly in~\cite{Schutz:1985km,Iyer:1986np,Konoplya:2003ii,Matyjasek:2017psv}.  

At high WKB orders (such as $N=6$ or $13$), the asymptotic expansion may display poor convergence due to alternating signs of successive terms. A simple and effective improvement is achieved by constructing a Padé rational approximant from the truncated WKB series, thereby extending its domain of validity. The Padé approximant of order $(m,n)$ is written as
\begin{equation}
P^{m}_{n}(x)=
\frac{a_0+a_1x+\cdots+a_mx^{m}}
{b_{0}+b_1x+\cdots+b_nx^{n}},
\label{eq:pade_approx}
\end{equation}
where $x$ symbolically represents the WKB expansion variable, for instance $x=(\omega^{2}-V_{0})/\sqrt{-2V_{0}^{\prime\prime}}$. In practice, Padé $(3,3)$ and versions yield quasinormal frequencies and GBFs in close agreement with fully numerical results obtained via continued-fraction or direct-integration methods~\cite{Matyjasek:2017psv, Konoplya:2019hlu}.

This combined WKB–Padé framework thus provides a rapid and reliable semi-analytic method for computing both QNMs and transmission coefficients. In the parameter range where the effective potential has a single smooth barrier, the results coincide with those from the time-domain evolution within a fraction of a percent, making the two approaches complementary and mutually validating.

\section{Quasinormal modes}\label{Quasinormal modes}

The quasinormal frequencies of a massive scalar field in the background of a Schwarzschild black hole immersed in a Hernquist dark-matter halo were computed using both the higher-order WKB approach with Pade approximants   (up to the 7th order) and the time-domain integration with Prony analysis.  The comparison between these methods demonstrates a very good level of consistency, with relative deviations in the real and imaginary parts of the frequencies typically below one percent.  This agreement confirms that the effective potential of the Hernquist halo, being a smooth single-barrier function, lies within the domain of validity of the WKB approximation.

From Tables~I–III, one observes that as the mass $\mu$ of the scalar field increases, the real part of the frequency $\omega_{\mathrm{R}}$ grows, while the imaginary part $\omega_{\mathrm{I}}$ decreases in magnitude.  This behavior, common to other black-hole configurations with massive perturbations, indicates that the oscillations become faster but more weakly damped as $\mu$ rises.  Extrapolation to large enough $\mu$ indicates that the damping rate should become very small, leading to the appearance of arbitrarily long-lived, quasi-resonant modes.  

The influence of the Hernquist profile enters through the characteristic scale $a_{0}$ and the total halo mass $M$.  For the astrophysically relevant hierarchy $M_{\mathrm{BH}}\!\ll\! M\!\ll\! a_{0}$, the changes in the quasinormal frequencies with varying $a_{0}$ are mild.  Increasing $a_{0}$ from $10^{2}$ to $10^{5}$ produces shifts in $\omega_{\mathrm{R}}$ and $\omega_{\mathrm{I}}$ about ten percents for all multipoles shown. 

The increase of the multipole number $\ell$ shifts the potential barrier to larger radii and increases its height, resulting in larger oscillation frequencies and shorter damping times.  For $\ell=0$, the modes exhibit the strongest sensitivity to $\mu$, while for $\ell=1$ and $\ell=2$ the variation becomes less pronounced, showing the expected convergence toward the eikonal regime.

The time-domain waveforms displayed in Figs.~\ref{fig:TD1} and~\ref{fig:TD2}  show the typical three-stage evolution: an initial transient, followed by an exponentially damped ringing and an oscillatory late-time tail with power-law envelope. The frequencies extracted from the Prony fit of the ringdown stage match the WKB results within the numerical precision reported in Tables~II–III, confirming the accuracy of the semi-analytic approach.

The overall behavior of the QNM spectra implies that, in the Hernquist halo, the presence of a massive scalar field can generate long-lived oscillations.  The gradual reduction of the damping rate with increasing field mass may enhance the contribution of such modes to the late-time tails of perturbations, while the dependence on the halo scale parameter ensures that these effects remain local to the black-hole vicinity.  Therefore, the obtained results support the robustness of black-hole quasinormal ringing against realistic dark-matter environments and clarify how the mass of the perturbed field, rather than the large-scale halo properties, governs the lifetime of the modes. These features are in agreement with observations for the other models of the galactic halo and for massless fields ~\cite{Konoplya:2021ube, Konoplya:2022hbl}. Altogether, this brings us to the conclusion that possible observations of minimally coupled waves must be safe against possible contamination by the galactic halo environment, unless the halo is exceptionally dense in the central region. However, the case of coupled fields deserves separate consideration \cite{Cardoso:2022whc}.

\begin{figure*}
\resizebox{\linewidth}{!}{\includegraphics{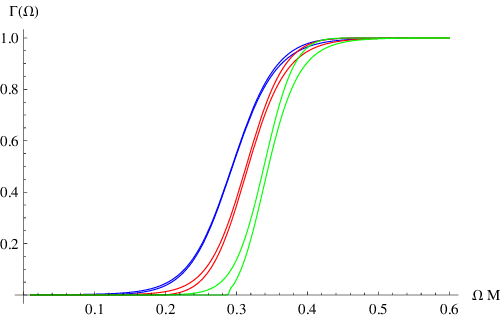}\includegraphics{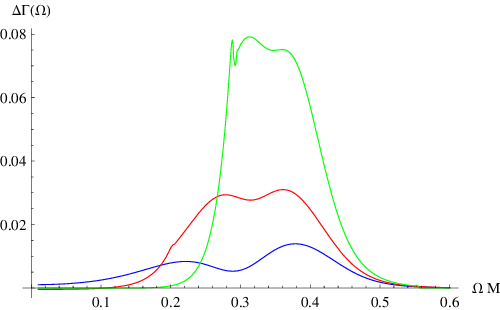}}
\caption{GBFs computed by the 6th order WKB method and by the correspondence with QNMs (left) and the difference between them (right) for $a_{0}=10^3$, $M=10 M_{BH}$, $\mu=0,0.2,0.3$ (blue, read, green - correspondingly), $\ell=1$, $M_{BH}=1$. }\label{fig:GBF1}
\end{figure*}

\begin{figure*}
\resizebox{\linewidth}{!}{\includegraphics{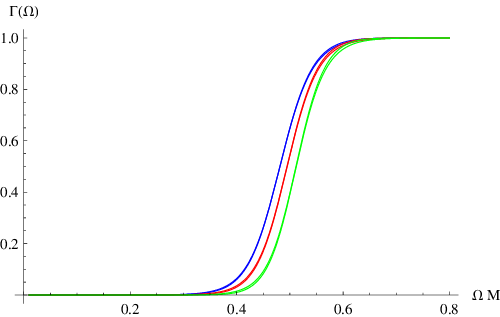}\includegraphics{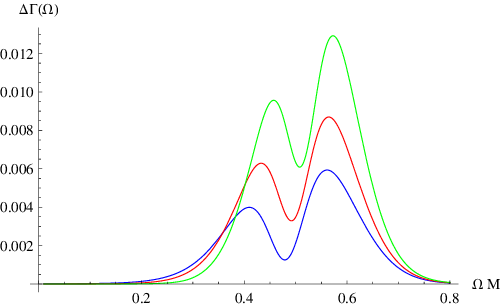}}
\caption{GBFs computed by the 6th order WKB method and by the correspondence with QNMs (left) and the difference between them (right) for $a_{0}=10^3$, $M=10 M_{BH}$, $\mu=0,0.2,0.3$ (blue, read, green - correspondingly), $\ell=2$, $M_{BH}=1$. }\label{fig:GBF2}
\end{figure*}

\begin{figure*}
\resizebox{\linewidth}{!}{\includegraphics{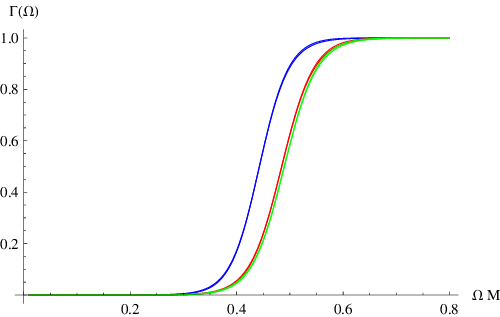}\includegraphics{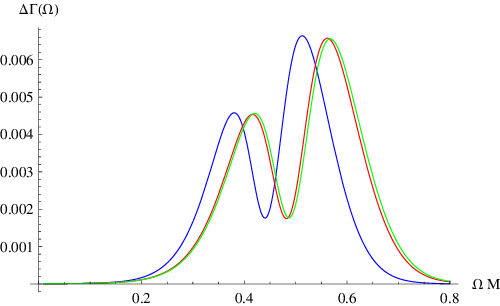}}
\caption{GBFs computed by the 6th order WKB method and by the correspondence with QNMs (left) and the difference between them (right) for $\mu=0.1$, $M=10 M_{BH}$, $a_{0}=100,10^{3},10^{5}$ (blue, read, green - correspondingly), $\ell=2$, $M_{BH}=1$. }\label{fig:GBF3}
\end{figure*}

\begin{figure*}
\resizebox{\linewidth}{!}{\includegraphics{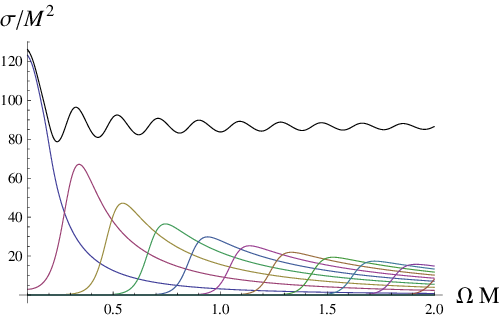}\includegraphics{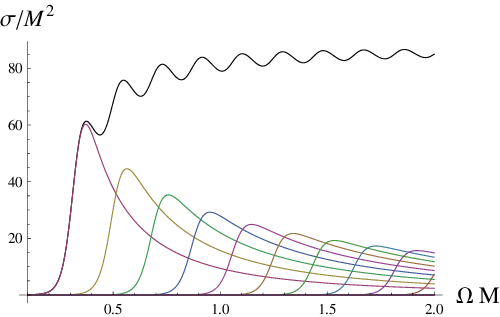}}
\caption{Absorption cross-section for the first 35 multipole numbers: $a_{0}=10^3$, $M=10 M_{BH}$, $\mu=0$ (left) and $\mu=0.25$ (right), $\ell=1$.}\label{fig:ACS}
\end{figure*}

\section{Grey-body factors and absorption cross-sections}\label{sec:GBF}

The higher-order WKB expansion, used previously for determining the QNM spectrum, can also be applied to estimate the transmission and reflection coefficients of the scattering problem. In the WKB framework, the transmission probability (i.e., the GBF) can be approximated by
\begin{equation}
\Gamma_{\mathrm{WKB}}(\omega)
=\left[1+\exp\!\bigl(2i\pi\,\mathcal{K}(\omega)\bigr)\right]^{-1},
\end{equation}
where
\begin{equation}
\mathcal{K}(\omega)
=\frac{\omega^{2}-V_{0}}{\sqrt{-2V_{0}^{\prime\prime}}}
-\sum_{k=2}^{N}\Lambda_{k},
\label{eq:WKBtransmission}
\end{equation}
with \(V_{0}\) and \(V_{0}^{\prime\prime}\) denoting the value and the second derivative of the effective potential at its maximum. This expression provides a reliable approximation of the true GBF $\Gamma(\omega)$ along the real frequency axis and has been widely employed in the literature for analyzing black-hole radiation and transmission properties~\cite{Konoplya:2023ahd, Dubinsky:2025ypj, Dubinsky:2025nxv, Dubinsky:2024vbn, Lutfuoglu:2025eik, Lutfuoglu:2025blw, Lutfuoglu:2025ldc, Bolokhov:2025lnt}.

The relation between the transmission coefficient and the QNM spectrum was first identified in~\cite{Konoplya:2024lir} and later examined and refined in a number of works~\cite{Bolokhov:2024otn, Skvortsova:2024msa, Malik:2024cgb, Malik:2025qnr, Malik:2025erb, Malik:2025dxn}. This so-called \emph{GBF/QNM correspondence} reveals that the poles of the transmission coefficient in the complex-frequency plane coincide with the quasinormal frequencies, at least in the eikonal regime where the WKB approximation is most accurate. At finite multipole numbers $\ell$, deviations appear, and the correspondence becomes approximate—as illustrated by the data obtained in the present study.

The correspondence fails in the same cases where the traditional relation between QNMs and null geodesics breaks down. This occurs, for instance, in higher-curvature theories of gravity where the effective potential at large $\ell$ no longer assumes the universal centrifugal form \(V(r) \sim g_{tt}\,\ell(\ell+1)/r^{2}\)~\cite{Konoplya:2017wot, Bolokhov:2023dxq, Konoplya:2017lhs}. A partial loss of correspondence is also known for asymptotically de Sitter geometries~\cite{Konoplya:2022gjp}, where the presence of a cosmological horizon alters the boundary conditions of the scattering problem.

A more refined relation valid for black holes and wormholes—including the leading and next-to-leading corrections beyond the eikonal regime—was derived in~\cite{Konoplya:2024lir, Bolokhov:2024otn}. In this formulation, the imaginary part of the function $\mathcal{K}$, which governs the transmission probability, can be expressed as \cite{Konoplya:2024lir,Bolokhov:2024otn} 
\begin{eqnarray}\nonumber 
&&\imo\K=\frac{\omega^2-\re{\omega_0}^2}{4\re{\omega_0}\im{\omega_0}}\Biggl(1+\frac{(\re{\omega_0}-\re{\omega_1})^2}{32\im{\omega_0}^2} \\\nonumber&&\qquad\qquad-\frac{3\im{\omega_0}-\im{\omega_1}}{24\im{\omega_0}}\Biggr) -\frac{\re{\omega_0}-\re{\omega_1}}{16\im{\omega_0}} \\\nonumber&& -\frac{(\omega^2-\re{\omega_0}^2)^2}{16\re{\omega_0}^3\im{\omega_0}}\left(1+\frac{\re{\omega_0}(\re{\omega_0}-\re{\omega_1})}{4\im{\omega_0}^2}\right) \\\nonumber&& +\frac{(\omega^2-\re{\omega_0}^2)^3}{32\re{\omega_0}^5\im{\omega_0}}\Biggl(1+\frac{\re{\omega_0}(\re{\omega_0}-\re{\omega_1})}{4\im{\omega_0}^2} \\\nonumber&&\qquad +\re{\omega_0}^2\Biggl(\frac{(\re{\omega_0}-\re{\omega_1})^2}{16\im{\omega_0}^4} \\&&\qquad\qquad -\frac{3\im{\omega_0}-\im{\omega_1}}{12\im{\omega_0}}\Biggr)\Biggr)+ \Order{\frac{1}{\ell^3}}. \label{eq:gbsecondorder} 
\end{eqnarray}
In this expression, $\omega$ denotes the real frequency of the emitted radiation, while $\omega_{0}$ and $\omega_{1}$ are the fundamental and first-overtone quasinormal frequencies, respectively. Equation~\eqref{eq:gbsecondorder} captures the leading corrections to the GBF/QNM correspondence and provides a high-precision tool for testing this link in numerical data.  

The absorption cross-section quantifies the effective area of a black hole (or any scattering center) that intercepts and absorbs incoming radiation. Physically, it represents the probability that an incoming wave of frequency $\omega$ and angular momentum $\ell$ will be transmitted through the potential barrier and enter the horizon, rather than being reflected back to infinity. For a field of frequency $\omega$, the total absorption cross-section is obtained by summing over all partial waves as
\begin{equation}
\sigma_{\mathrm{abs}}(\omega)
=\frac{\pi}{\omega^{2}}
\sum_{\ell=0}^{\infty}(2\ell+1)\,\Gamma_{\ell}(\omega),
\label{eq:sigma_abs}
\end{equation}
where $\Gamma_{\ell}(\omega)$ is the grey-body (transmission) coefficient for the corresponding multipole. In the low-frequency limit, $\sigma_{\mathrm{abs}}$ approaches the area of the event horizon, $\sigma_{\mathrm{abs}}\to A_{\mathrm{H}}$, while at high frequencies it oscillates around the geometric-optics capture cross-section associated with the unstable photon orbit. Thus, $\sigma_{\mathrm{abs}}(\omega)$ provides a bridge between the wave and particle descriptions of radiation, encapsulating both quantum and classical aspects of black-hole interaction with external fields \cite{Futterman:1988ni}.

The frequency-dependent GBFs of the massive scalar field were evaluated using the WKB method for several values of the halo parameters and field mass. The general behavior, displayed in Figs.~\ref{fig:GBF1}, ~\ref{fig:GBF2} and~\ref{fig:GBF3}, follows the standard pattern for black-hole scattering potentials: at low frequencies the transmission probability is strongly suppressed, while for $\omega$ exceeding the potential barrier the GBF approaches unity. The peak structure of $\Gamma_{\ell}(\omega)$ becomes broader and lower as the field mass increases, reflecting the reduction of the barrier height and the emergence of a potential plateau at large $r$. Consequently, the halo and mass terms act in opposite directions—while the halo parameters $M$ and $a_{0}$ slightly enhance transmission at fixed $\omega$, the finite field mass tends to suppress it. These trends are consistent for all multipole numbers shown, confirming that the influence of the Hernquist halo on the scattering properties remains perturbative compared to the dominant near-horizon geometry.

The total absorption cross-section, computed from the partial-wave sum in equation~\eqref{eq:sigma_abs} and illustrated in Fig.~\ref{fig:ACS}, exhibits the expected smooth transition between the low- and high-frequency regimes. For small $\omega$, $\sigma_{\mathrm{abs}}$ tends to the horizon area, indicating that the near-horizon contribution dominates the absorption process. At intermediate frequencies, oscillations appear around the geometric-optics limit due to interference between reflected and transmitted waves, a feature typical for spherically symmetric black holes. As $\mu$ increases, these oscillations are damped, and the overall magnitude of $\sigma_{\mathrm{abs}}$ decreases, in agreement with the suppression of $\Gamma_{\ell}(\omega)$ at low frequencies. Varying the halo parameters $M$ and $a_{0}$ leads only to marginal shifts in the absorption curves, confirming that the large-scale dark-matter distribution has a mild impact on the energy transmission efficiency.  

Overall, the obtained GBFs and absorption cross-sections indicate that the Hernquist halo modifies the scattering characteristics of massive fields only slightly, without altering their qualitative behavior. The radiation emitted by the black hole remains largely governed by the local geometry near the horizon, while the dark-matter environment introduces only small corrections at higher orders in $M/a_{0}$.

\section{Conclusions}\label{sec:conc}

In this work, we have analyzed QNMs, GBFs, and absorption cross-sections of a massive scalar field propagating in the spacetime of a Schwarzschild black hole surrounded by a Hernquist dark-matter halo. The model provides an analytic and asymptotically flat description of a spherically symmetric halo and allows for a consistent study of the interplay between the local black-hole geometry and the large-scale dark-matter distribution.

The QNM spectra were computed by means of the higher-order WKB approximation and confirmed through time-domain integration. Both methods yielded mutually consistent results, showing that the WKB–Padé scheme remains accurate for the smooth single-barrier potentials produced by the Hernquist profile. The real parts of the quasinormal frequencies increase with the effective mass of the scalar field, while the imaginary parts decrease in magnitude, indicating the appearance of long-lived, quasi-resonant modes. Variations in the halo parameters, such as its total mass and scale radius, were found to cause only relatively small shifts in the frequencies, demonstrating that the ringdown signal is largely insensitive to the extended dark-matter environment.

The analysis of the GBFs and absorption cross-sections confirmed that the main scattering features of the field are governed by the near-horizon geometry. The transmission probability decreases at low frequencies and approaches unity in the high-frequency limit, following the typical pattern of black-hole scattering. Increasing the field mass lowers the GBFs and the overall absorption efficiency, while modifications of the halo parameters have only a marginal effect. The absorption cross-section smoothly interpolates between the horizon area at low energies and the geometric-optics capture value at high frequencies, displaying weak oscillations in the intermediate regime.

\begin{acknowledgments}
The author is grateful to Excellence Project PrF UHK 2205/2025-2026.
\end{acknowledgments}

\bibliography{bibliography}

\end{document}